\documentclass[10pt,notitlepage,showpacs,pra,twocolumn]{revtex4-1}
\usepackage{amsfonts}
\usepackage{amsmath}
\usepackage{amssymb}
\usepackage{graphicx}
\usepackage{float}
\usepackage[toc,page,header]{appendix}
\usepackage{color}
\usepackage{multirow}

\begin{document}

\title{Deterministic facilitated excitation of the weakly-driven atom in heteronuclear Rydberg atom pairs beyond antiblockade}
\author{Han Wang, Dongmin Yu, Rui Li and Jing Qian$^{\dagger}$ }
\affiliation{State Key Laboratory of Precision Spectroscopy, Quantum Institute for Light and Atoms, Department of Physics, School of Physics and Electronic Science, East China
Normal University, Shanghai 200062, China}

\begin{abstract}
Due to the intrinsic strong blockaded interaction shifting the energy level of Rydberg state, the steady Rydberg probability may be substantially restrained to a low level, especially for atoms suffering from weak drivings. We report an exotic excitation facilitation using strong blockaded energy shifting the double excitation state, leading to the weakly-driven atom deterministically excited by a significantly improved fraction even at far off-resonant regimes. This phenomenon is attributed to the effective induction of a reduced detuning to the weakly-driven atom by optimizing parameters of an auxiliary strongly-driven atom, arising a modified large probability preserved for a broadening range of detunings. The influence from breaking blockade condition on the excitation probability is also discussed. In contrast to previous approaches, our excitation facilitation mechanism does not rely on antiblockade effect to the compensation of Rydberg energy, serving as a fresh way to a deterministic single-atom excitation of the target atom in a pair of heteronuclear Rydberg atoms, where a strong interatomic interaction is required.
\end{abstract}
\email{jqian1982@gmail.com}
\maketitle
\preprint{}

\section{Introduction}

The possibility to all-optical detection of highly-excited state population with high-precision has opened a new avenue to the investigation of different excitation facilitation mechanisms \cite{Schonleber14,Anderson16,Bhowmick19}. If detecting a single two-level atom of weak driving $\omega<|\delta|$ ($\omega$ denotes the excitation Rabi frequency and $\delta$ is the detuning to the upper level), the steady-state excitation probability of the upper-level can be well characterized by a Lorentzian function of detuning $\delta$ \cite{Goldschmidt16}
\begin{equation}
P_A = \frac{4\omega^2}{1+8\omega^2+4\delta^2}
\label{PA}
\end{equation}
with the frequency unit $\gamma$. According to the formula (\ref{PA}) it is intuitive that the direct way accessible to a facilitated excitation can be decreasing the detuning $\delta$ via ac Stark shift \cite{Zanjani07} or using a high-power laser $\omega$, allowing it towards its steady saturation value $P_A\to0.5$ on resonance $\delta\ll\omega$ \cite{Olmos14}. However if $\omega<|\delta|$ is assumed, the excitation probability $P_A $ is poor and rapidly decreases as $|\delta|$ increases. For more atoms with strong interatomic interactions ({\it e.g.} superatom) the presence of dephasing will cause the average excitation probability significantly higher, approaching 1.0 with the increase of atomic number within the superatom \cite{Honer11,Petrosyan13,Petrosyan132}. 
Other studies resort to an effective two-level system deducing from a three-level configuration on the assumption of a large detuning to the middle state \cite{Marzoli94,Wu97,Dudin12}, enabling the steady saturation value reaching 1.0 via tunable two-mode laser fields in an adiabatic process \cite{Bergmann98}. A transient non-steady higher probability can also be obtained via coherent excitation with special pulsed lasers, applying for constructing multifarious quantum gate devices \cite{Isenhower10}. Nevertheless, in most cases due to the complexity of realistic atomic structures, facilitating dynamic excitation for weakly-driven atoms remains an elusive and open question for current experiments.

When the uppermost level is replaced by a Rydberg state, a traditional facilitation phenomenon known as Rydberg excitation antiblockade, predicted by Ates {\it etc.}\cite{Ates07,Ates072} and experimentally verified by Weidem\"{u}ller group in $^{87}$Rb atoms \cite{Amthor10}, occurs when the interaction induced energetic shift can be well compensated by the laser detuning to the Rydberg level, triggering many interesting applications then. 
For instance it enables the implementation of an accurate one-step quantum logic gate \cite{Su17} or a maximal entanglement \cite{Shao17,Zhao17}, facilitating excitation dynamics even in thermal atoms beyond the frozen limit \cite{Li13,Marcuzzi17,Kara18} or in anisotropic interacting systems due to partial resonance \cite{Qian13}. Besides, some nontrivial facilitation approaches, such as relying on inhomogeneous broadening on blue-detuned atoms with attractive interactions \cite{Letscher17}, avoided crossings between collective many-body states \cite{Garttner14,Chai17}, or seeded avalanche dynamics in off-resonant Rydberg atoms \cite{Simonelli16}, have been proposed, providing more prospects for observing strongly correlated excitation growth in ultracold Rydberg gases \cite{Sanchez18,Wu17,Thaicharoen18}.
All these facts are essentially based on a tunable detuning overcoming the shift caused by the Rydberg-Rydberg interaction, facilitating the population on a double Rydberg state. In other words, antiblockade can only give an excitation growth for the Rydberg pair state, while it fails to deterministic excitation of a target atom.

In the present work, our purpose is providing a simplified method for how to facilitate the excitation probability of a weakly-driven target atom via blockade effect. To our knowledge, dressing atoms to Rydberg levels will arise strong interatomic interactions blocking the excitation of Rydberg pair state. In a strong blockade scenario, the single Rydberg states $|gr\rangle$ or $|rg\rangle$ (for atomic number $N=2$) would be prepared with same possibility, despite that the pumping laser of each atom is weak or strong \cite{Urban09,Pritchard10}. With Rydberg blockade effect, researchers have achieved single-atom excitation via preparing single-atom source from unknown numbers of atoms \cite{Beterov11,Petrosyan14,Petrosyan15}. For homonuclear atoms the probability to arbitrary single excitation state is same due to the simultaneous excitation by same driving pulses \cite{Heidemann07,Wilk10,Beguin13}.
However for two heteronuclear atoms this property may be used to facilitate the excitation of a single atom deterministically. This virtue of using heteronuclear atoms also lies in different resonant frequencies of the two atoms that allows the addressing and manipulating of them individually in experiment, arising a strong heteronuclear Rydberg blockade.

The core concept lies in employing an auxiliary heteronuclear strongly-driven atom that blockadedly interacts with the target weakly-driven atom, resulting in a reduced detuning $|\delta^{\prime}|<|\delta|$ with respect to the target atom. The effective detuning $\delta^{\prime}$ has absorbed the influences from the Rabi frequency $\Omega$ and the detuning $\Delta$ of the strongly-driven atom, so an optimization for the values of $\Omega$ and $\Delta$ are required. It is observed that the modified excitation probability has an anomalous excitation growth by 2 orders of magnitude compared to the case containing purely one atom, especially robustly at far off-resonance. Our findings well agree with recent experimental results where a large spectral broadening for the steady Rydberg excitation probability due to resonant dipole-dipole interaction was reported \cite{Goldschmidt16}. 
In the end, we discuss the effect of antiblockade on the facilitated single-atom probability by reducing the strength of Rydberg interactions (breaking the blockade condition) and present a summary of all parameters required under the reasonable experimental conditions. 
Our proposal not only provides a possible way for efficiently exciting weakly-driven atoms with more complex structures, but also  opens a new route to deterministic excitation of target atoms from heteronuclear Rydberg-blockaded ensembles.

\section{effective $V$-type model}

\begin{figure}
\includegraphics[width=3.4in,height=3.0in]{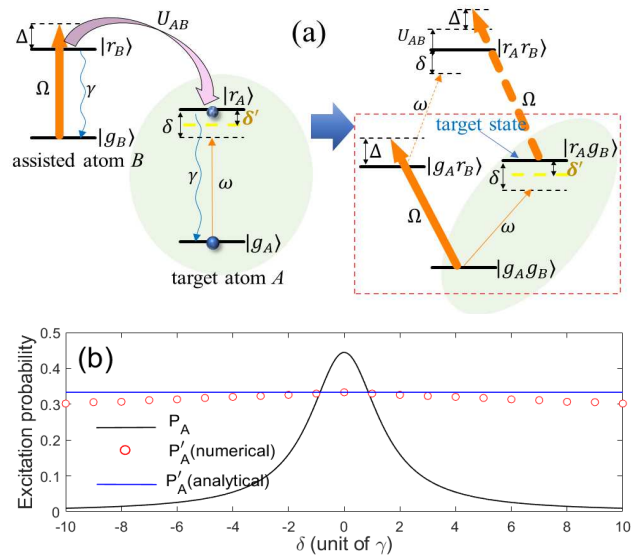}
\caption{(color online). (a) With an auxiliary strongly-driven atom $B$, the target weakly-driven atom $A$ can be off-resonantly excited to the Rydberg state $|r_A\rangle$ with a modified enhanced probability. An equivalent picture on the right side shows, under an assisted atom $B$, the expected detuning between the ground $|g_Ag_B\rangle$ and the target states $|r_Ag_B\rangle$ can be described by an effective reduced detuning $|\delta^{\prime}|<|\delta|$ if both $\Delta$ and $\Omega$ are optimized. Also the double Rydberg state $|r_Ar_B\rangle$ is assumed to be safely discarded due to its unoccupancy in the strong blockade environment. (b) The singly-excited Rydberg probability $P_A$(black solid, from Eq.(\ref{PA})) and the modified probability $P_A^{\prime}$ (population of state $|r_Ag_B\rangle$, red circles), versus the change of $\delta$. An approximated analytical value $P_A^{\prime}\approx0.33$ (blue solid) that does not depend on $\delta$ under the conditions of $\Omega\gg\omega$ and $|\Delta|\gg|\delta|$, is shown. Enlarging $|\delta|$ arises a bigger deviation between analytical and numerical results due to the breakup of condition $|\Delta|\gg|\delta|$. Here $\omega=1$ and $\gamma$ is the frequency unit throughout the paper.}
\label{modelfig}
\end{figure}

The description of a realistic scheme is represented in Figure \ref{modelfig} where the target atom $A$ consisting of ground state $|g_A\rangle$ and Rydberg state $|r_A\rangle$, is off-resonantly driven by the laser Rabi frequency $\omega$ and detuning $\delta$, satisfying $|\delta|>\omega$(weak driving). An auxiliary heteronuclear atom $B$ excited by a strongly-driven Rabi frequency $\Omega$ with a detuning $\Delta$ to the Rydberg level $|r_B\rangle$, can induce a significant interatomic interaction $U_{AB}$ shifting the energy level of $|r_Ar_B\rangle$. We first consider the strong blockaded condition with $U_{AB}\gg|\delta|, |\Delta|$ that can entirely block the excitation of $|r_Ar_B\rangle$, essentially different from the antiblockade mechanism that uses appropriate $\delta, \Delta$ to compensate the shift $U_{AB}$ for a higher Rydberg probability. In this case the scheme can be treated as an effective three-level $V$ system, facilitating more comprehensive discussions. It is shown that by suitably adjusting $\Omega$ and $\Delta$ a reduced detuning $\delta^{\prime}$ to the target state $|r_Ag_B\rangle$ can be realized, enabling an anomalous spectra broadening for the excitation probability of weakly-driven atom.

To demonstrate the physics we begin with the effective Hamiltonian for the simplified $V$-type structure,
\begin{equation}
\hat{\mathcal{H}}_{V}=-\delta\hat{\sigma}_{22}+\Delta\hat{\sigma}_{33}+\frac{\omega}{2}(\hat{\sigma}_{12}+\hat{\sigma}_{12}^{\dagger})+\frac{\Omega}{2}(\hat{\sigma}_{13}+\hat{\sigma}_{13}^{\dagger}) 
\label{Hammv}
\end{equation}
with operator $\hat{\sigma}_{ij}=|i\rangle\langle j|$($i,j\in\{1,2,3,4\}$), states $|1\rangle=|g_Ag_B\rangle$, $|2\rangle=|r_Ag_B\rangle$, $|3\rangle=|g_Ar_B\rangle$, $|4\rangle=|r_Ar_B\rangle$ accordingly (note that $|4\rangle$ is omitted in this section). The time evolution of state population can be solved following the Lindblad master equation: $\dot{\hat{\rho}}(t)=i[\hat{\mathcal{H}}_{V},\hat{\rho}]+\hat{\mathcal{L}}[\hat{\rho}]$ where the density matrix elements $\rho_{ij}$ are governed by the optical Bloch equations where we ignore the dephasing rate of atomic polarization here,
\begin{eqnarray}
\dot{\rho}_{11} &=& \gamma(\rho_{22}+\rho_{33})+\frac{i\omega}{2}(\rho_{12}-\rho_{21})+\frac{i\Omega}{2}(\rho_{13}-\rho_{31})\nonumber\\
\dot{\rho}_{22} &=& -\gamma\rho_{22}-\frac{i\omega}{2}(\rho_{12}-\rho_{21}) \nonumber\\
\dot{\rho}_{33} &=& -\gamma\rho_{33}-\frac{i\Omega}{2}(\rho_{13}-\rho_{31}) \label{boeq} \\
\dot{\rho}_{12} &=& -(\frac{\gamma}{2}+i\delta)\rho_{12}+\frac{i\omega}{2}(\rho_{11}-\rho_{22})-\frac{i\Omega}{2}\rho_{32} \nonumber\\
\dot{\rho}_{13} &=& -(\frac{\gamma}{2}-i\Delta)\rho_{13}-\frac{i\omega}{2}\rho_{23}+\frac{i\Omega}{2}(\rho_{11}-\rho_{33}) \nonumber\\
\dot{\rho}_{23} &=& -(\gamma-i\delta-i\Delta)\rho_{23} - \frac{i\omega}{2}\rho_{13}+\frac{i\Omega}{2}\rho_{21} \nonumber
\end{eqnarray}

Directly solving Eqs.(\ref{boeq}) in the steady state condition ($\dot{\rho}_{ij}=0$) gives rise to the steady populations. Here the population of state $|r_Ag_B\rangle$[$=|2\rangle$] presenting the modified excitation probability of the target atom, can be defined as $P_A^{\prime}=\langle 2|\hat{\rho}(t)|2\rangle$, taking a similar form as Eq.(\ref{PA}),
\begin{equation}
P_A^{\prime} = \frac{4\omega^2}{1+8\omega^2+4[\delta^{\prime}(\Delta,\Omega,\delta,\omega)]^2}
\label{PAp}
\end{equation}
In deriving Eq.(\ref{PAp}) we have assumed that $\gamma$($\gamma^{-1}$) is the frequency(time) unit. Explicitly, the modified detuning $\delta^{\prime}$ becomes a more complex function now which can be further simplified under approximations $\Omega\gg\omega$ and $|\Delta|\gg|\delta|$, letting it depend on $\Omega$ and $\Delta$ only,
\begin{widetext}
\begin{equation}
\delta^{\prime}(\Delta,\Omega)=\frac{48\Delta^4+4\Delta^2(15+20\Omega^2+4\Omega^4)+8\Omega^6+35\Omega^4+44\Omega^2+12}{64\Delta^4+16\Delta^2(5+4\Omega^2)+4(2+\Omega^2)^2}
\label{PAan}
\end{equation}
\end{widetext}

The resulting approximated value $P_{A}^{\prime}$[$\approx 0.33$] is kept to be unvaried with respect to $\delta$ when both $\Delta$ and $\Omega$ are optimal [see Sec. 3], as shown in Fig.\ref{modelfig}(b)(blue solid). However the exact function $\delta^{\prime}(\omega,\delta,\Omega,\Delta)$ is a more complex form that has to be solved numerically. 

In Figure \ref{modelfig}(b) we numerically simulate the modified steady excitation probability $P_A^{\prime}$ to the target atom versus $\delta$. The Lorentzian function $P_A$  is comparably presented in the same picture. We observe that $P_A$ decreases dramatically with $|\delta|$, reaching as low as 0.01 for $|\delta|=10$. Surprisingly, with the influence of another auxiliary strongly-driven atom, the modified probability $P_{A}^{\prime}$ benefits from a significant improvement at far off-resonance. The numerical results identify that $P_{A}^{\prime}$ having a broadening linewidth only suffers from a tiny decrease ($\Delta P_{A}^{\prime}<0.05$) by increasing $|\delta|$ from 0 to 10, in contrast to a dramatic decrease of  $P_A$ ($\Delta P_{A}>0.44$) for the presence of unique atom $A$. The modified excitation probability $P_{A}^{\prime}$ anomalously benefits from 

(i) a high excitation probability even at far-off resonance $|\delta|\gg\omega$; 

(ii) a relatively steady Rydberg excitation rate regardless of the absolute value of $\delta$.

This excitation facilitation mechanism can also be understood in the dressed-state picture with the assistance of $\Omega\gg\omega$ and $\Delta\gg\delta$. The subspace of two strongly-coupled states $\{|1\rangle,|3\rangle\}$ can be described by the dressed states $|\pm\rangle = (\mathcal{N}_{\pm})^{-1}(\Delta\pm\sqrt{\Delta^2+\Omega^2})|1\rangle-\Omega|3\rangle$ with $\mathcal{N}_{\pm}$ the normalization coefficient, arising two transition paths $|+\rangle\to|2\rangle$ and $|-\rangle\to|2\rangle$ respectively. The resulting quantum interference between them can arise a significant improved excitation to the upper state $|2\rangle$ roughly at $\delta = (\sqrt{\Delta^2+\Omega^2}-\Delta)/2$ by avoided level crossing.

\begin{figure}
\includegraphics[width=3.4in,height=1.7in]{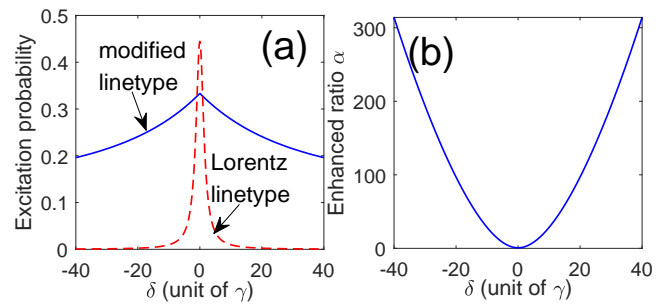}
\caption{(color online). (a) The modified excitation probability $P_{A}^{\prime}$ (blue solid) broadening for a wider range of  detunings $\delta\in[-40,40]$. For comparison the Lorentz-linetype probability $P_{A}$ (red dashed) is also presented. (b) The enhanced ratio defined by $\alpha=P_{A}^{\prime}/P_A$ versus $\delta$.}
\label{wider}
\end{figure}

An extensive study for a broad range of $\delta$ shows that the modified probability $P_{A}^{\prime}$ also has a small decreasing with $\delta$. As shown in Fig.\ref{wider}(a) by increasing $|\delta|$ to as large as $40$, $P_{A}^{\prime}$ (blue solid) is reduced to 0.20 accompanied by a continuous increasing for the enhanced ratio $\alpha = P_{A}^{\prime}/P_{A}$ which reaches as high as 300 at $\delta=\pm40$ [see Fig.\ref{wider}(b)]. On the contrary, the single excitation probability $P_A$ (red dashed) functioning as a Lorentz linetype suffers from a dramatical decrease, presenting a very poor output at far-off resonance. 

\section{parameter optimization of the auxiliary atom}

\begin{figure}
\includegraphics[width=3.3in,height=3.3in]{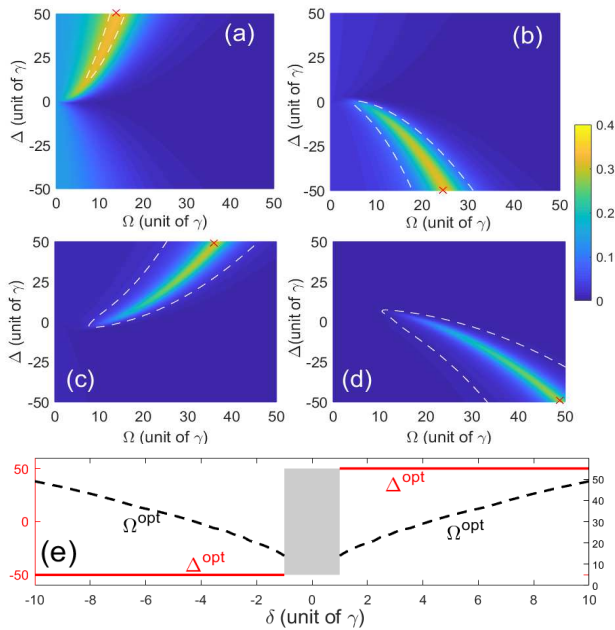}
\caption{(color online). The modified excitation probability $P_{A}^{\prime}$ versus the variations of $\Omega$ and $\Delta$ from the auxiliary atom $B$. (a-d) $\delta=(1.0, -3.0, 6.0, -10)$ accordingly and $\omega=1$. The white dashed curve represents the boundaries where $P_A = P_A^{\prime}$ meaning no enhancement. Here the strong-blockaded condition $U_{AB}=1000$ is used. (e) The optimal values $\Omega^{opt}$ and $\Delta^{opt}$ vs $\delta$ whose positions are denoted by red crosses in (a-d). The shadow region stands for near-resonance case where $|\delta|>\omega$ is not satisfied, arising no enhanced excitation for the target atom, see Fig.\ref{modelfig}(b).}
\label{enhanced}
\end{figure}

The primary results provided in previous section depend on an optimization of $\Delta$ and $\Omega$ of an auxiliary strongly-driven atom $B$, enabling a reduced detuning $|\delta^{\prime}|\ll|\delta|$. For this motivation we freely tune $\Delta\in[-50,50]$ and $\Omega\in[0,50]$ in order to search for an optimal combination of values $\Delta$ and $\Omega$ that can realize a modified and optimal probability output. Noting that if the search range changes the modified probability $P_A^{\prime}$ will have a slight change.

For instance we choose $\delta=(1.0, -3.0, 6.0, -10)$, fixing the tunable regimes of $\Delta$ and $\Omega$, then we plot the modified probability $P_A^{\prime}$ as a function of both $\Omega$ and $\Delta$ in Fig.\ref{enhanced}(a-d). In each picture it is interesting to observe a crescent-like regime where $P_A^{\prime}>P_A$ can be met (the white dashed curve denotes the boundary where $P_A^{\prime}=P_A$). The labelings marked by red crosses point to the position of maximal value. For a deep analysis, we observe that for an excitation facilitation of the weakly-driven atom, one requires an optimized and large value of $\Delta$ having same sign with respect to $\delta$, which can partly cancel each other for an improved excitation [see Eq.(\ref{Hamful})]. Here the sign of $\Delta^{opt}$ is determined by $\delta$ and its absolute value keeps $|\Delta^{opt}|=50$ (large detuning, $|\Delta^{opt}|\gg|\delta|$) for a big suppression to the excitation of atom $B$, as plotted in Fig.\ref{enhanced}(e)[red solid].

Figure \ref{enhanced}(a-d) also show the optimal laser Rabi frequency $\Omega^{opt}$ of the strongly-driven atom continuously grows with $|\delta|$, intuitively coinciding with the facts that a stronger laser is required if detuning is off-resonance. Figure \ref{enhanced}(e) represents a conclusive plot of the exact optimal values of $\Omega^{opt}$ and $\Delta^{opt}$ as a function of $\delta$, calculated numerically. In the regime $|\delta|>\omega$, for the auxiliary atom, $\Delta^{opt}$ has an in-phase sign with $\delta$ however the absolute value of which is constant regardless of $\delta$. On the other hand, $\Omega^{opt}$ reveals a linear increase with $|\delta|$. That means, in a real implementation in order to anomalously excite the weakly-driven atom $A$, one only needs to take a scanning for an optimal $\Omega^{opt}$ that depends on the absolute value of $|\delta|$, yet $\Delta^{opt}$ is highly insensitive to $|\delta|$. In other words for a given $\Delta$ by increasing $\Omega$ there always exists an optimal $\Omega^{opt}$ that can enable a modified enhanced excitation, offering a more flexible selection of parameters for the scheme implementation.

\section{breaking strong blockade}

In the following we study an extensive case if the strong blockade condition breaks, {\it i.e.} the double excitation state $|r_Ar_B\rangle$[=$|4\rangle$] does affect, arising a diamond-type configuration [see right panel of Fig.\ref{modelfig}(a)], with the Hamiltonian given by
\begin{equation}
\hat{\mathcal{H}}_{\diamond} = \hat{\mathcal{H}}_{V}+\hat{\mathcal{H}}_{\Lambda}
\end{equation}
where the first term presenting the effective three-level model given in Eq.(\ref{Hammv}), and the second term $\hat{\mathcal{H}}_{\Lambda}$ for interactions with double excitation state, is described by
\begin{equation}
\hat{\mathcal{H}}_{\Lambda}=\frac{\omega}{2}(\hat{\sigma}_{34}+\hat{\sigma}_{34}^{\dagger})+\frac{\Omega}{2}(\hat{\sigma}_{24}+\hat{\sigma}_{24}^{\dagger}) +(U_{AB}-\delta+\Delta)\hat{\sigma}_{44}
\label{Hamful}
\end{equation}

\begin{figure}
\includegraphics[width=3.4in,height=2.3in]{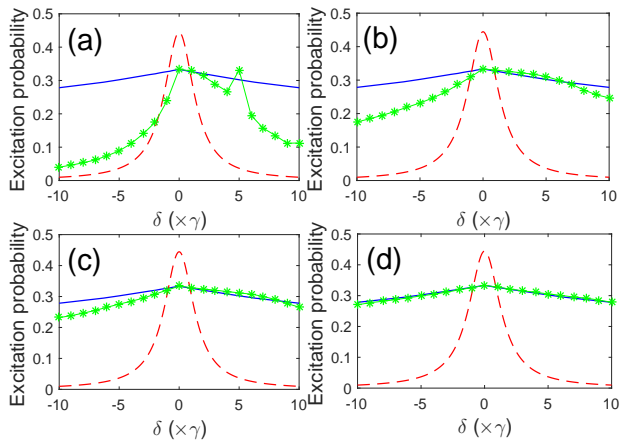}
\caption{(color online). The single-atom excitation $P_A$ (red dashed, Lorentz-type), the modified excitation $P_A^{\prime}$ (blue solid) under strong blockade with $U_{AB}=1000$ and the modified excitation $P_A^{\prime\prime}$ under weaker interactions $U_{AB}=(10,50,100,500)$ (green curve with stars) in (a-d), versus detuning $\delta$.   }
\label{res}
\end{figure}

Presently numerical calculations should be solved with the whole Hamiltonian $\hat{\mathcal{H}}_{\diamond}$ and the density operator $\hat{\rho}(t)$ being a $4\times4$ matrix then, where the detected quantity turns to be $P_A^{\prime\prime} =\langle 2|\hat{\rho}(t)|2\rangle+\langle 4|\hat{\rho}(t)|4\rangle$ since state $|4\rangle$ does affect now. By considering $U_{AB}=(10,50,100,500)$ we investigate the excitation probability $P_A^{\prime\prime}$ versus detuning $\delta$ in Figure \ref{res}(a-d), denoted by green curve with stars. We note that when the interaction strength $U_{AB}$ is small it could perfectly compensate the difference between detunings $\delta$ and $\Delta$, arising an enhanced excitation fraction if the critical antiblockade condition [{\it i.e.} $U_{AB}=\delta-\Delta$] is met. Remarkably, at $\delta=5.0$ there exists an abnormal  peak probability with $U_{AB}=10,\Delta=-5$, caused by antiblockade [see Fig.\ref{res}(a)]; whereas, increasing $U_{AB}$ will lead to the breakup of this critical condition. 

In Figure \ref{res}(b-c), owing to the use of a larger $U_{AB}$, tuning $\delta$ in the regime of $[-10,10]$ is not sufficient for perfectly compensate this energy shift $U_{AB}$ since $\Delta$ is set to be within the regime of $[-50,50]$. Hence an asymmetric population distribution $P_A^{\prime\prime}$ is found due to the partial compensation on the positive detunings. Finally as $U_{AB}$ is tuned to be 500 that is extremely larger than all detunings, just like $U_{AB}=1000$ used in previous sections, no clear compensation effect is observable. To this end, the double excitation state $|4\rangle$ is far off-resonance, giving rise to a symmetric distribution for the single excitation probability $P_A^{\prime\prime}$ with respect to the detuning $\delta$.

For comparison, $P_A$ (red dashed) and $P_A^{\prime}$ (blue solid) are presented in same pictures [see Fig.\ref{res}(a-d)]. In general, the modified probability $P_A^{\prime}$ has a dominant excitation at $|\delta|>\omega$ (weak-driving). However as decreasing the strength of Rydberg interactions, the modified probability $P_A^{\prime\prime}$ affected by the double excitation state would cause a decreasing and asymmetric population distribution because the assisted strongly-driven atom $B$ is also comparably excited under weaker interactions, see relevant parameters summarized in Table \rm{I} of section \rm{V}.

\section{Realistic parameters required}

 For real experimental implementation we make numerical calculations with reasonable parameters of $^{87}Rb$ and $^{85}Rb$ atom pairs, the decay rate (frequency unit) of which is estimated to be $\gamma/2\pi=36.5$kHz for the energy levels with principal quantum number $n>60$ \cite{Beterov09}. In addition, we assume the dipole-dipole coefficient $C_3/2\pi=5.24$GHz$\mu m^3$ and the average spacing $R$ between two heteronuclear atoms can be tuned within a few micrometers, leading to $U_{AB}=C_3/R^3$. The heteronuclear Rydberg blockade is experimentally accessible by increasing the blockade shifted energy to as high as $ (0.1\sim1.0)$ GHz at zero temperature \cite{Zeng17}. Relevant numerical results under experimental parameters are summarized in Table \ref{ta}.
 
The target atom $A$ is assumed to be triggered by an unvaried laser with Rabi frequency $\omega/2\pi=0.0365$MHz, off-resonantly coupled to the upper state by a large detuning value $\delta = 10\omega$. For single atom $A$ the excitation fraction is very poor, {\it i.e.} $P_A=0.0098$. Surprisingly, when introducing another auxiliary atom $B$ bringing considerable atom-atom interactions between them, the excitation of atom $A$ can obtain a significant enhancement, reaching as large as $\approx0.3124$ if full blockade emerges [at $R\approx5.24\mu$m $U_{AB}\gg\omega,\Omega$]. Meanwhile we note that the excitation of auxiliary atom $B$ sustains a low level $\sim 0.01$, becoming one order of magnitude weaker than the excitation of atom $A$ in the case of full blockade. That fact can be attributed to the interaction-induced reduction of detuning, arising an increased excitation deterministically for the target atom $A$. In other words, the full blockade facilitates the excitation probability of only one atom no matter its coupling strength is strong or weak. To quantitatively compare with the  result from dressed-state picture that the enhanced excitation is predicted to be roughly at $\delta/2\pi = (\sqrt{\Delta^2+\Omega^2}-\Delta)/2\approx0.463$MHz, which has a small derivation from the optimal numerical value $\delta/2\pi=0.365$MHz as shown in Table \ref{ta}, owing to the inadequate blockaded condition.

\begin{widetext}

\begin{table}[tbp]
\begin{tabular}{c|c|c|c|c|c|c|c|c}
\hline\hline
  \multirow{2}*{different cases}& \multirow{2}*{R/$\mu$m} & \multirow{2}*{$U_{AB}$/$2\pi$(MHz)} & \multicolumn{3}{|c|}{target atom $A$} & \multicolumn{3}{|c}{auxiliary atom $B$} \\  \cline{4-9}
 & & & $\omega$/$2\pi$(MHz) & 
 $\delta$/$2\pi$(MHz) & $P_A^{(\prime)}$ & $\Omega$/$2\pi$(MHz) & $\Delta$/$2\pi$(MHz) & $P_B^{(\prime)}$ \\ 
\hline
\multirow{4}*{reaching blockade} & $\backslash$ & $\backslash$  & 0.0365 & 0.3650 &0.0098  & $\backslash$ & $\backslash$  & $\backslash$  \\ \cline{2-9}
 & 24.30 & 0.3650 & 0.0365 & 0.3650 & 0.0031&  4.40  & 10.00 & 0.0461 \\ \cline{2-9}
  &14.21 & 1.825 & 0.0365 & 0.3650 & 0.0037 &  4.40  & 10.00 & 0.0440 \\ \cline{2-9}
 & 5.24 & 36.50 & 0.0365 & 0.3650 & 0.3124 &  4.40  & 10.00 & 0.0390 \\ 
  \hline
  antiblockade & 24.30  & 0.3650 & 0.0365 & 0.1825 & 0.3312  & 1.825 & -0.1825  & 0.4952 \\
  \hline
 \multirow{2}*{optimizing blockade} & 5.24 & 36.5  & 0.0365 & 0.3650 & 0.0507  & 3.90  & 10.00  & 0.0526   \\ \cline{2-9}
 & 5.24 & 36.5 & 0.0365 & 0.3650 & 0.0450 &  4.90   & 10.00 & 0.0349   \\ \cline{2-9}
\hline\hline
 \end{tabular}%
\caption{A full numerical comparison using reasonable experimental parameters including the inter-atomic spacing $R$, the Rydberg interaction $U_{AB}$, laser Rabi frequencies $\omega$, $\Omega$ and energy-level detunings $\delta$, $\Delta$ from a pair of heteronuclear Rydberg atoms $^{85}Rb-^{87}Rb$ considered in our scheme. For the case of reaching blockade via decreasing $R$ to 5.24$\mu$m, the modified excitation probability $P_A^{(\prime)}$ [=0.3124] turns to be much larger than that of the assisted atom $B$ [=0.0390] by one order of magnitude. The antiblockade case represents a simultaneously facilitated excitation of two atoms when $U_{AB}=\delta-\Delta$ is met. In the optimizing-blockade case, the importance of optimizing the Rabi frequency $\Omega$ is verified by the dramatic reduction of $P_A^{(\prime)}$ as adding a small deviation [$\pm0.5$MHz] to the optimal value $\Omega^{opt}$[=$2\pi\times$4.4MHz].}%
\label{ta}
\end{table}


\end{widetext}

From Table \ref{ta} the importance of optimizing parameters $\Omega$ of atom $B$ is also clearly seen. When $\Omega$ is tuned to be away from its optimal value [$\Omega^{opt}\approx2\pi\times4.4$MHz], here $\Omega/2\pi=3.9$MHz or $4.9$MHz, we observe that $P_A^{\prime}$ will have a dramatic decrease by one orders of magnitude.
In addition it also shows one special case of antiblockade where $U_{AB}=\delta-\Delta$ is exactly met. Note that here atoms $A$ and $B$ both represent high and comparable excitation probabilities [$P_A^{\prime}=0.3312$, $P_B^{\prime}=0.4952$] by the compensation of Rydberg shift, which is insufficient for the motivation of deterministic excitation of the weakly-driven atom in our protocol. Therefore the scheme proposed here relying on Rydberg blockade effect, is robust to excite one target atom while excitation of the other atom in heteronuclear atom pair can be efficiently suppressed owing to its far-off-resonance. Quite different from the previous schemes using antiblockade mechanism, here blockade is required as an active ingredient for excitation facilitation, whereas antiblockade has exchanged its role to be a detrimental ingredient.

\section{conclusion}

To conclude, we investigate an anomalous excitation facilitation in heteronuclear Rydberg atom pairs with strong Rydberg-Rydberg interactions. The preparation is that one atom is weakly-driven and the other auxiliary atom is strongly-driven. Intuitively the strongly-driven atom is easier to be excited. However we identify that with the presence of the auxiliary strongly-driven atom
there exists a regime that the excitation fraction of the weakly-driven atom can be dramatically enhanced to be one order of magnitude larger than the other auxiliary atom. The reason is attributed to the fact that the auxiliary atom induces an reduced detuning with respect to the target atom, anomalously arising an excitation facilitation accompanied by an increased linewidth to it at off-resonant regimes $|\delta|>\omega$. A simplified $V$-type three-level model is analytically discussed in detail, revealing that the essence for this facilitation is blockadedly shifting the double Rydberg level, facilitating the excitation of single atom. Meanwhile the importance of optimizing parameters for the auxiliary atom is also stressed.

Additionally our work confirms that, under experimental conditions, we can achieve a deterministic facilitation only to the target atom, the auxiliary atom can be safely stayed in the ground state without excitation although it is strongly-driven. Our proposal is an interesting extension to the usual facilitation mechanism based on antiblockade, which may be realizable in current experiments. Furthermore it may lead to unexpected behaviors to the community of heteronuclear Rydberg atoms, offering new applications for simulations of many-body excitation dynamics between hetero-species atoms.

For prospect, with the consideration of more surrounding atoms, if the interactions between $N$ assisted atoms are strong enough enabling a single collective excitation between the ground state $|G_B\rangle = |g_1g_2...g_N\rangle_B$ and the one-atom excited state $|R_B\rangle =\frac{1}{\sqrt{N}} \sum_{i=1}^{n=N}|g_1...r_i...g_N\rangle_B$, the collective oscillation is characterized by an enhanced Rabi frequency $\sqrt{N}\Omega$ between $|G_B\rangle$ and $|R_B\rangle$, constituting a two-level superatom \cite{Stanojevic09}. To this end, the persistence of this facilitated excitation effect of the target atom $A$ with the help of a blockaded superatom ensemble can be predicted, and a detailed analysis will be left for future studies.

\acknowledgements

This work was supported by the NSFC under Grants No. 11474094, No. 11104076, the ``Fundamental Research Funds for the Central Universities'', the Academic Competence Funds for the outstanding doctoral students under YBNLTS2019-023, the
Specialized Research Fund for the Doctoral Program of Higher Education No.
20110076120004.

\end{document}